\documentclass[11pt,a4paper]{article}

\usepackage[margin=2.5cm]{geometry}
\usepackage{graphicx}
\usepackage{amsmath}
\usepackage{amssymb}
\usepackage{booktabs}
\usepackage{hyperref}
\usepackage{url}
\usepackage{caption}
\usepackage{placeins}   

\graphicspath{{figures/}}

\makeatletter
\renewcommand{\@openbib@code}{%
  \setlength{\topsep}{0pt}%
  \setlength{\itemsep}{1pt plus 0.3ex}%
  \setlength{\parsep}{0pt}%
  \setlength{\partopsep}{0pt}%
}
\makeatother

\title{Differentiated Services:\\
       an \emph{Experimental vs. Simulated} Case Study\thanks{%
  \copyright{} 2002 IEEE. Personal use of this material is permitted.
  Permission from IEEE must be obtained for all other uses, including
  reprinting/republishing this material for advertising or promotional
  purposes, creating new collective works, for resale or redistribution
  to servers or lists, or reuse of any copyrighted component of this
  work in other works. The version of record is published by the IEEE at
  \url{https://doi.org/10.1109/ISCC.2002.1021705}. This is the author's
  accepted manuscript (preprint), typeset by the author and released
  for open archival.\par\medskip
  This work has been partially supported by the Telecommunications
  Laboratory of Lappeenranta University of Technology (Finland) as
  part of the ILIAS Project.\par\medskip
  \noindent\rule{\linewidth}{0.3pt}\par\smallskip
  {\small
  \textit{Cite as:} S.~Andreozzi, ``Differentiated Services: an
  Experimental vs.\ Simulated Case Study,'' in \textit{Proc.\ IEEE
  ISCC 2002}, pp.~383--390.\\
  Version of record (IEEE):~\url{https://doi.org/10.1109/ISCC.2002.1021705}\\
  Author-prepared preprint (Zenodo):~\url{https://doi.org/10.5281/zenodo.19665017}
  \quad{\footnotesize(this PDF snapshot:~\url{https://doi.org/10.5281/zenodo.19665018})}\\
  Source thesis, 2001 MSc (Zenodo):~\url{https://doi.org/10.5281/zenodo.19662899}\\
  Companion software archive, ns-2 (Zenodo):~\url{https://doi.org/10.5281/zenodo.19665019}%
  \par}}}

\author{Sergio Andreozzi\\[2pt]
        Istituto Nazionale di Fisica Nucleare -- CNAF\\
        Viale Berti Pichat, 6/2 -- 40127 Bologna, Italy\\
        \texttt{sergio.andreozzi@cnaf.infn.it}}

\date{}

\begin{document}

\nocite{ciscollq,ns2home,nsdiffserv,tftant,andreozziNs,bennett1996,%
        ipdvdraft,rfc2679,golestani1994,goyal1996,rfc2598,rfc3246,%
        parekh1993,rfc2475,stiliadis1996,ferrari2000}

\maketitle

\begin{abstract}
This paper aims to provide a proof of concept of the accuracy of
simulations for advanced networking study. The particular target
technology is the Differentiated Services (DiffServ) architecture.
The method has been to apply experimental activities conducted in a
real network to a simulation environment, to gather the same
performance parameters and to compare results.

A worthy re-engineering of the DiffServ module of the deployed
software program has been carried out and significant contribution
have been made to overcome the encountered limitations and to enrich
its modeling capabilities. Final results give useful suggestions for a
more critical approach to simulations targeted for advanced networking
study.
\end{abstract}

\section{Introduction}

In the field of advanced networking, simulations are usually conducted
to understand how new technologies affect the network activity. Since
the behavior of real systems can be duplicated through hardware and
software, the knowledge of both systems implementation design and
simulator modeling capabilities are an important requirement. The
understanding of what can be properly replicated is a necessary
building block for a valuable research work.

The DiffServ architecture is the prominent resource allocation scheme
defined by IETF~\cite{rfc2475} to provide scalable services
differentiation in the Internet. The TF-TANT task force~\cite{tftant}
conducted several experiments of this architecture in a real European
large-scale test bed. The objective, here referred, was to study the
behavior of delay and jitter sensitive traffic in an IP
DiffServ-enabled network using different mechanisms and parameter
settings~\cite{ferrari2000}. This paper replicates both TF-TANT test
bed and experiments in a simulation environment (the Network Simulator
or NS~\cite{ns2home}). A proof of concept of how faithful to the
reality the simulation activity can be is given.

Through all this work, a worthy re-engineering of the Network
Simulator (NS) DiffServ module has been done to overcome the
encountered limitations. Moreover, new mechanisms suitable for the
DiffServ architecture implementation have been provided.

\section{Experimental vs. Simulated}

\begin{figure}[t]
  \centering
  \includegraphics[width=0.6\linewidth]{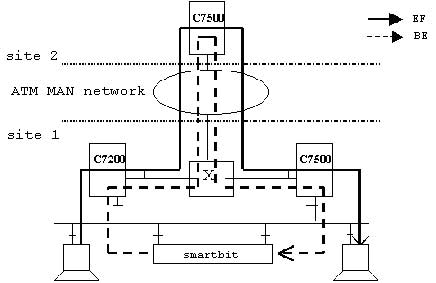}
  \caption{Experimental testbed~\cite{ferrari2000}}
  \label{fig:1}
\end{figure}

In the TF-TANT DiffServ experiments~\cite{ferrari2000}, a test bed
with an edge DiffServ-enabled router was set up (Figure~\ref{fig:1}).
A Metropolitan Area Network (MAN), based on ATM technology, provides
a Constant Bit Rate (CBR) connection of 2~Mbps between two edge
routers, both connected to a Local Area Network (LAN). This LAN is
based on Fast Ethernet technology and active devices (switches) are
used. The Cisco 7200 router is the edge router DiffServ-enabled,
while the two Cisco 7500 routers are set up to behave as
non-congested FIFO devices. Packets crossing egress router
interfaces, after the queuing and scheduling stages, experience
additional buffering needed to be transmitted. This extra buffering
stage can be modeled as a FIFO queue cascaded to the queuing and
scheduling systems and it is usually called Transmission queue (TX
queue).

The test bed porting to the simulation environment is based on the
following considerations (also summarized in Table~\ref{tab:1}):

\begin{itemize}
\item in the real scenario the LAN is built using switches, the line
rate is 100 Mbps and the conveyed traffic rate is 2 Mbps (plus MAC
protocol overhead). It is meaningful to assert that, under these
conditions, the shared media property of this technology does not
affect in a sensitive manner the conveyed traffic. Thereby,
connections between hosts and routers can be modeled as
point-to-point links.

\item in the real scenario both traffic generators and sinks for
monitored traffic are in the same device (Smartbit card). This is to
avoid the synchronization error due to different clocks. Since the
simulator program runs in a stand-alone machine in which a general
clock is available to the whole system, sources and destinations can
be split into different hosts.

\item in the real scenario routers have a Transmission queue (TX
queue) cascaded to the queuing and scheduling system; this queue is
not present in the simulated router model. In this, the scheduler
steers packets directly to the link. The decision is to not solve
this limitation because tests are targeted for the Expedited
Forwarding (EF) aggregate in a situation of congestion. Therefore
this aggregate is assigned for a service rate equal to or greater
than the arrival rate~\cite{rfc3246} and the TX queue contributes to
the queuing delay for the time to be emptied.

\item in the real scenario the MAN is based on ATM technology.
Routers are connected with a 2~Mbps CBR Virtual Path. These are
modeled as a point-to-point link whose line rate is 2~Mbps.
\end{itemize}

\begin{figure}[t]
  \centering
  \includegraphics[width=0.8\linewidth]{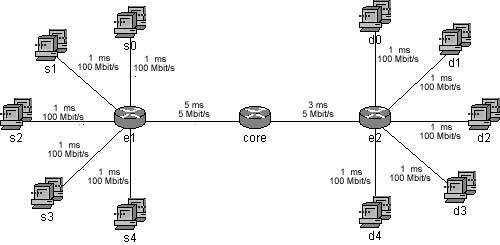}
  \caption{Simulated testbed}
  \label{fig:2}
\end{figure}

\begin{table}[t]
  \centering
  \caption{Real to simulated porting summary}
  \label{tab:1}
  \begin{tabular}{@{}lll@{}}
    \toprule
    Feature            & Real                         & Simulated          \\
    \midrule
    LAN technology     & switch                       & p2p links          \\
                       & fast Ethernet                & 100 Mbps,          \\
                       &                              & 1 ms delay         \\
    MAN technology     & ATM,                         & p2p links          \\
                       & 2 Mbps CBR PVC               & 2 Mbps CBR         \\
    Router             & with TX queue                & none               \\
    egress interface   & cascaded to scheduler        &                    \\
    \bottomrule
  \end{tabular}
\end{table}

Using these assumptions, the test bed has been ported to the
simulator environment (Figure~\ref{fig:2}). E1 is the DiffServ-enabled
router, S0--S4 are the transmitters and D0--D4 are the receivers. In
the following section, a brief analysis of the NS DiffServ module
characteristics~\cite{nsdiffserv} against the test requirements is
given and contributions are described.

\section{The NS DiffServ module and new contributions}

In the current NS (release 2.1b8a), the official DiffServ module is a
porting of a Nortel Networks contribution. Characteristics of this
module have been analyzed and matched against the TF-TANT experiments
needs. The analysis pointed out several lacks that can be classified
in two main categories:

\begin{itemize}
\item functional blocks:
  \begin{itemize}
  \item available schedulers are only Priority Queuing (PQ), Round
    Robin (RR), Weighted Round Robin (WRR) and Weighted Interleaved
    Round Robin (WIRR)~\cite{nsdiffserv}; the need is for the
    Weighted Fair Queuing (WFQ) scheduler, Cisco implementation
    (Self-Clocked Fair Queuing~\cite{golestani1994}).
  \item classification, marking and metering actions are coupled in
    one command; there is the need for the creation of aggregates
    from different flows through marking and for the metering on an
    aggregate basis.
  \item a drop-out-profile dropper, useful for the definition of an
    arrival rate limiter (e.g. token bucket meter + in-out marker +
    out-profile dropper) on an aggregate basis is not present.
  \end{itemize}
\item performance parameters monitoring features: One-Way Delay (OWD)
  and IP Delay Variation (IPDV) can be computed only maintaining
  trace files; drawbacks are that these files can be very large and
  need to be processed.
\end{itemize}

A consistent part of this work has been to provide an improved NS
DiffServ module~\cite{andreozziNs} that solves the above limitations.
Firstly, new schedulers have been implemented. In particular, an
abstract class has been defined for scheduler algorithms and existing
schedulers have been moved to derived classes. Using this design,
several new schedulers have been easily added: Packet-by-packet
Generalized Processor Sharing (PGPS)~\cite{parekh1993}, Worst-Case
Weighted Fair Queueing plus (WF$^2$Q+)~\cite{bennett1996}, Start-Time
Fair Queuing (SFQ)~\cite{goyal1996}, Low Latency Queuing
(LLQ)~\cite{ciscollq} and Self-Clocked Fair Queuing
(SCFQ)~\cite{golestani1994}. Secondly, new performance parameters
have been added. In particular both end-to-end One-Way Delay
(OWD)~\cite{rfc2679} and end-to-end IP Delay Variation
(IPDV)~\cite{ipdvdraft} can be now computed at simulation time,
without the need for a storing and post-processing of trace files.
The instantaneous value, the average and the frequency distribution
can be easily gathered. Finally, the script level command interface
has been revised to give access to the new features.

In the following sections both experiments and simulations are
described and compared. The selected tests are taken from a TF-TANT
technical report~\cite{ferrari2000}. These tests focus on DiffServ
support for delay and jitter sensitive traffic, as follows:

\begin{itemize}
\item Test A. analysis of bandwidth over-provisioning and packet size
  impacts on OWD using WFQ
\item Test B. impact of Best Effort (BE) traffic packet size on EF
  OWD and IPDV using PQ
\item Test C. WFQ vs. PQ, comparison of average OWD, average IPDV,
  OWD frequency distribution and IPDV frequency distribution for
  different EF packet sizes
\end{itemize}

For tests comparison, the general rule is that, if simulations
provide accurate results, no experimental test diagram is showed.
This can be checked in~\cite{ferrari2000}.

\section{Test A: bandwidth over-provisioning and packet size impacts
  on OWD using WFQ}

The WFQ (SCFQ implementation) is a suitable scheduler for high speed
networks for which, low computational cost, fair bandwidth sharing
among several queues, excessive bandwidth distribution and delay
bound guarantees are fundamental properties. Each queue is assigned
with a weight that corresponds to the dedicated amount of service
time that is proportional to the sum of all weights. The purpose of
this test is to understand how much over-provisioning of service time
affects the queuing time experienced by EF packets. The parameter
took into consideration is the Service-To-Arrival-Ratio ($STAR$)
defined as:
\begin{equation}
  S_r \cdot l_r = A_r \cdot STAR
\end{equation}
in which $S_r$ is the service rate expressed as a fraction of the
line rate, $l_r$ is the link rate and $A_r$ is the EF arrival rate.
The impact of $STAR$ parameter on OWD is analyzed.

\subsection{Experimental test}

In the TF-TANT experiment, two services are configured
(Table~\ref{tab:2}). The Premium Service for the EF traffic and the
Background Service for the BE traffic. Tests are repeated for
several EF packets size and for different values for the STAR
parameter. The end-to-end OWD for EF packets is monitored and a
diagram is provided.

\begin{table}[t]
  \centering
  \caption{Traffic generator and services setting for router Cisco 7200}
  \label{tab:2}
  \begin{tabular}{@{}llllll@{}}
    \toprule
    Service    & PHB & Load   & Frame    & Protocol & Service \\
               &     & (Kbps) & size     &          & rate    \\
    \midrule
    Premium    & EF  & 300    & variable & UDP      & $S_r \cdot l_r$ \\
    Background & BE  & 2000   & 1000     & UDP      & remaining       \\
    \bottomrule
  \end{tabular}
\end{table}

\subsection{Simulated test}

In the simulated test, the WFQ scheduler and two queues are
configured in the e1$\rightarrow$core egress interface
(Figure~\ref{fig:2}) to support both Premium and Background
aggregates. The assumption for the EF aggregate is to have traffic
entering the EF queue at a constant rate of 300 Kbps. For this
reason, the EF aggregate is generated by a unique CBR source and is
policed with a token bucket meter (Committed Information Rate or
CIR=300Kbps, Committed Burst Size or CBS=1 packet) and a
drop-out-profile dropper. To avoid synchronization problems due to
the deterministic start time, background traffic is generated with
several CBR sources whose rate is chosen from a uniform random
distribution in the range [10~Kbps, 100Kbps], while the starting
time is chosen from a uniform random distribution in the range
[0s, 5s]. The BE packet size is always 1000 bytes. A single
simulation conducted for a 200-second period is characterized by a
certain EF packet size and a certain STAR parameter according to the
TF-TANT settings. The average OWD is computed on the whole period.

\subsection{Comparison}

\begin{figure}[t]
  \centering
  \includegraphics[width=0.85\linewidth]{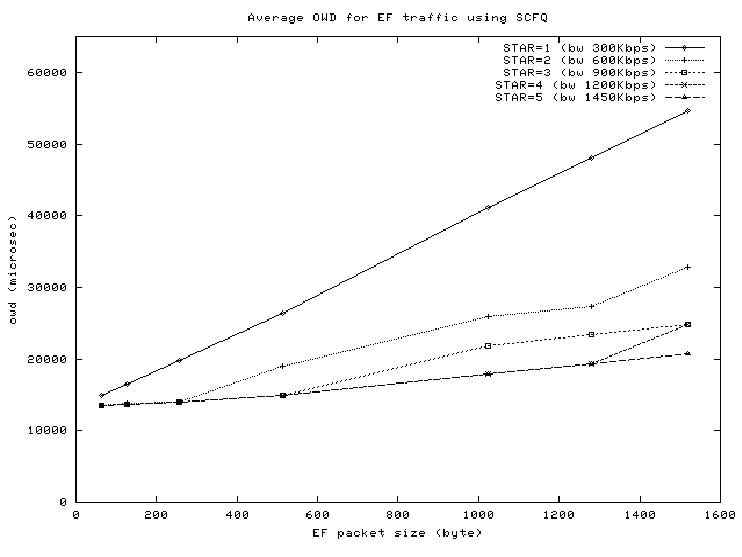}
  \caption{EF average OWD vs. packet size for WFQ scheduler -- simulation}
  \label{fig:3}
\end{figure}

Both real and simulated tests show that the increment of the STAR
parameter significantly reduces the OWD, in particular for large
packet sizes (Figure~\ref{fig:3}). The gain is relevant until the
STAR value equals 4. In~\cite{ferrari2000} it is asserted that IPDV
is not sensitive to this parameter; this property has been positively
verified through simulation. A discrepancy between the two tests is
perceivable for STAR parameter greater than 1 and short packet sizes.
The simulation experiment shows a stable OWD, while the experiment in
the real network shows a raising OWD.

\section{Test B: impact of BE packet size on EF OWD and IPDV using PQ}

Using the Priority Queuing (PQ) scheduler, packets at the head of a
certain queue have always the right to be served over packets
belonging to lower priority queues. But preemption is made only
during scheduling action, therefore on-going lower priority packet
transmission can delay the service time of a certain packet. The
more lower priority packets are big, the more they can affect the
queuing time of higher priority packets. This test aims to study
this phenomena.

\subsection{Experimental test}

In the TF-TANT experiment, two services are configured
(Table~\ref{tab:3}): The Premium Service for the EF traffic and the
Background Service for the BE traffic. Tests are repeated for two
different EF packet sizes and for different BE packet sizes. Both
average OWD and average IPDV for EF packet size are monitored for
different BE packet sizes.

\begin{table}[t]
  \centering
  \caption{Traffic generator and services setting for router Cisco 7200}
  \label{tab:3}
  \begin{tabular}{@{}lllll@{}}
    \toprule
    Service    & PHB & Load (Kbps) & Frame size & Protocol \\
    \midrule
    Premium    & EF  & 300         & 128, 1024            & UDP \\
    Background & BE  & 2000        & $\in$[100, 1450]     & UDP \\
    \bottomrule
  \end{tabular}
\end{table}

\subsection{Simulated test}

In the simulated test, the PQ scheduler and two queues are
configured in the e1$\rightarrow$core egress interface
(Figure~\ref{fig:2}) to support both EF and BE aggregates. The
assumption for the EF aggregate is to have traffic entering the EF
queue at a constant rate of 300 Kbps. For this reason, the EF
aggregate is generated by a unique CBR source and is policed with a
token bucket meter (CIR=300Kbps, CBS=1 packet) and a
drop-out-profile dropper. To avoid synchronization problems due to
the deterministic start time, background traffic is generated with
several sources whose rate is chosen from a uniform random
distribution in the range [10 Kbps, 100Kbps], while the starting
time is chosen from a uniform random distribution in the range
[0s, 5s]. A single simulation conducted for a 200-second period is
characterized by a certain EF packets size and a certain BE packet
size according to the TF-TANT settings. The average OWD and average
IPDV are computed on the whole period.

\subsection{Comparison}

\begin{figure}[t]
  \centering
  \includegraphics[width=0.75\linewidth]{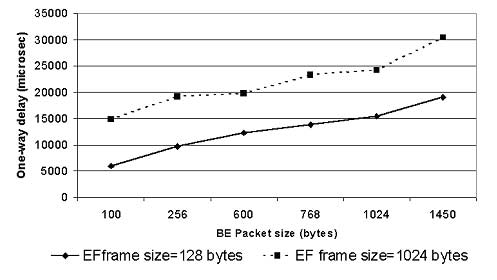}
  \caption{EF average OWD vs. BE packet size for PQ scheduler --
           experiment~\cite{ferrari2000}}
  \label{fig:4}
\end{figure}

\begin{figure}[t]
  \centering
  \includegraphics[width=0.85\linewidth]{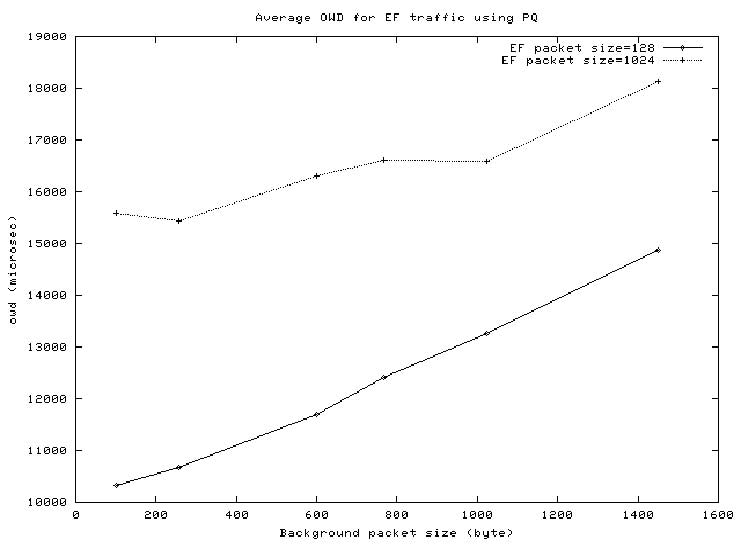}
  \caption{EF average OWD vs. BE packet size for PQ scheduler -- simulation}
  \label{fig:5}
\end{figure}

In both tests, the average OWD increases with BE packet size
(Figures~\ref{fig:4} and~\ref{fig:5}), but in the simulated network,
this parameter raises less than in the real network. A more worthy
comparison could take place using OWD normalized to the optimal one
in both networks.

\begin{figure}[t]
  \centering
  \includegraphics[width=0.75\linewidth]{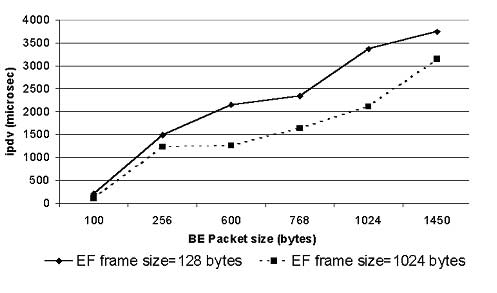}
  \caption{EF average IPDV vs. BE packet size for PQ scheduler --
           experiment~\cite{ferrari2000}}
  \label{fig:6}
\end{figure}

\begin{figure}[t]
  \centering
  \includegraphics[width=0.85\linewidth]{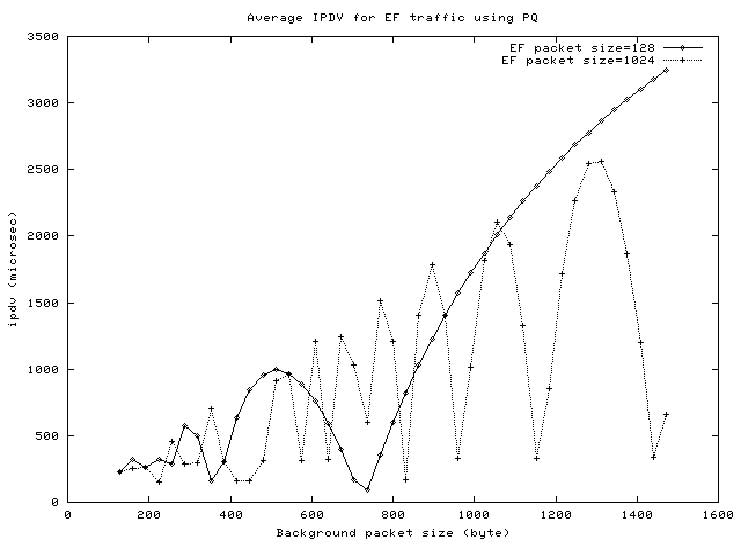}
  \caption{EF average IPDV vs. BE packet size for PQ scheduler -- simulation}
  \label{fig:7}
\end{figure}

Referring to IPDV (Figures~\ref{fig:6} and~\ref{fig:7}), results
present more discrepancies. While in the real network, the average
IPDV for EF packets presents a raising trend, in the simulated test
an irregular behavior was monitored. The showed diagram has been
generated using a wider set of BE packet size values to better
describe the phenomena.

For 1024-byte EF packet size, a saw toothed trend is evidenced. It
is interesting to notice that, repeating several times the simulation
and considering that background traffic sources always have different
start times and different rates, both max and min in the curves were
unchanged.

\section{Test C: WFQ vs. PQ for OWD and IPDV parameters}

The PQ scheduler is easy to implement and strongly recommended for
delay sensitive traffic. Excess bandwidth is shared among backlogged
services in a priority manner. Drawbacks are the need of an accurate
policy to avoid starvation of lower priority services and the
difficulty in giving delay bound guarantees to all services.

Conversely, the WFQ scheduler provides links bandwidth sharing
property, excess bandwidth distribution based on service rate and
delay bound properties to all services. This test aims to explore
the behavior of EF OWD and IPDV for different EF packet sizes when
using these two schedulers.

\subsection{Experimental test}

In the TF-TANT experiment, two services are configured
(Table~\ref{tab:4}). The Premium Service for the EF traffic and the
Background Service for the BE traffic.

\begin{table}[t]
  \centering
  \caption{Traffic generator and services setting for router Cisco 7200}
  \label{tab:4}
  \begin{tabular}{@{}lllll@{}}
    \toprule
    Service    & PHB & (Kbps)    & Frame size         & Protocol \\
    \midrule
    Premium    & EF  & 300       & $\in$[64, 1518]    & UDP \\
    Background & BE  & $>$~2000  & variable           & UDP \\
    \bottomrule
  \end{tabular}
\end{table}

Tests are repeated for different EF packet sizes and for both
schedulers. The background traffic is composed by four independent
streams with different packet size. Both OWD and IPDV for EF packet
size are monitored, in particular, average and frequency distribution
are computed.

\subsection{Simulated test}

In the simulated test, two queues are configured in the
e1$\rightarrow$core egress interface (Figure~\ref{fig:2}) to support
both EF and BE aggregates. The assumption for the EF aggregate is to
have traffic entering the EF queue at a constant rate of 300 Kbps.
For this reason, the EF aggregate is generated by a unique CBR
source and is policed with a token bucket meter (CIR=300Kbps,
CBS=1 packet) and a drop-out-profile dropper. To avoid
synchronization problems due to the deterministic start time, the
background traffic is generated with several sources whose starting
time is chosen from a uniform random distribution in the range
[0s, 5s]. Moreover, each source has a 100 Kbps rate, but the packet
size varies from 64 to 1472 bytes (64-byte increment) for a total
of 23 flows. This decision is intended to create the more various
background traffic. No detailed TF-TANT settings for traffic
background were available.

A single simulation conducted for a 200-second period is
characterized by a certain EF packet size and a certain scheduler
according to the TF-TANT settings. The average OWD, the average
IPDV, the OWD frequency distribution and the IPDV frequency
distribution are computed on the whole period.

\subsection{Comparison}

\begin{figure}[t]
  \centering
  \includegraphics[width=0.85\linewidth]{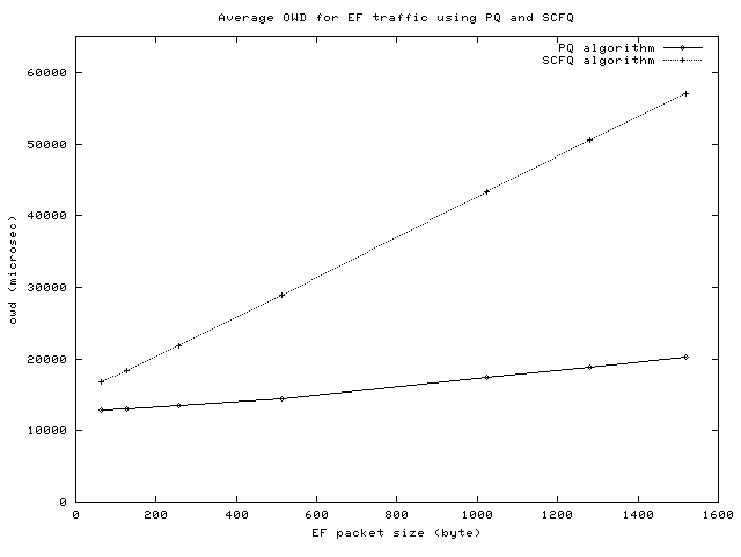}
  \caption{Average OWD with WFQ and PQ for different EF packet sizes --
           simulation}
  \label{fig:8}
\end{figure}

Firstly, average OWD parameter is compared. Both diagrams show that
the PQ scheduler performs better than WFQ scheduler
(Figure~\ref{fig:8}). The PQ gain increases with the EF packet size.
As explained in TF-TANT experiments~\cite{ferrari2000}, the PQ
queuing delay is mainly introduced by the ongoing transmission of a
lower priority packet, which needs to be terminated before an EF
packet is scheduled for transmission. On the other hand, with WFQ an
additional delay source has to be taken into account: the time needed
to wait until all packets with a smaller forwarding time in the
queuing system are scheduler for transmission~\cite{golestani1994}.

\begin{figure}[t]
  \centering
  \includegraphics[width=0.85\linewidth]{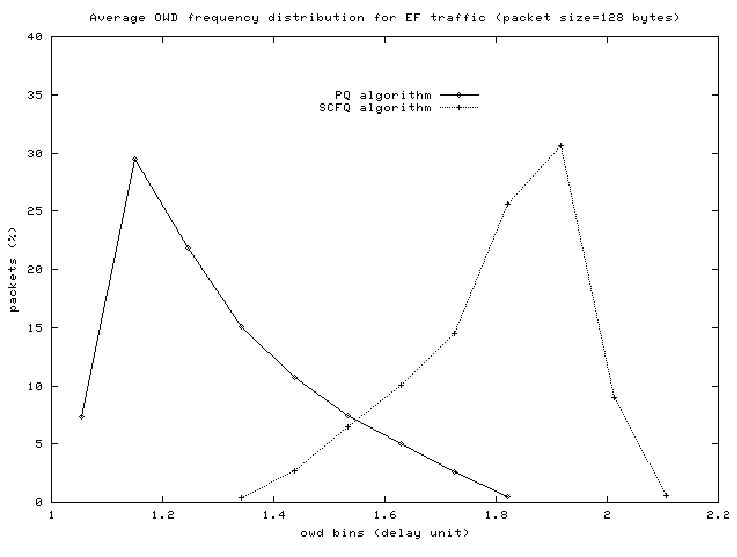}
  \caption{Frequency distributed OWD with WFQ and PQ for 128-byte EF
           packet size -- simulation}
  \label{fig:9}
\end{figure}

\begin{figure}[t]
  \centering
  \includegraphics[width=0.85\linewidth]{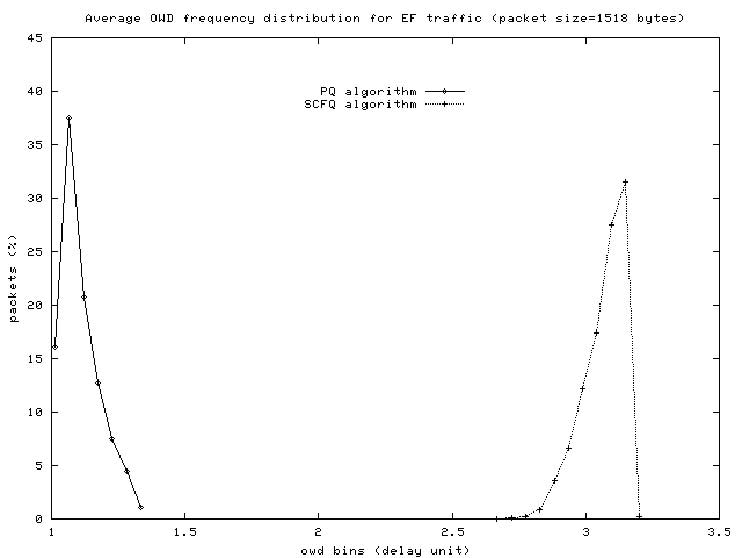}
  \caption{Frequency distributed OWD with WFQ and PQ for 1518-byte EF
           packet size -- simulation}
  \label{fig:10}
\end{figure}

Figure~\ref{fig:9} shows OWD frequency distribution for 128-byte EF
packet size, while Figure~\ref{fig:10} shows OWD frequency
distribution for 1518-byte EF packet size. In each diagram, the
delay unit is equivalent to the minimum OWD experienced by EF
packets and, for simulations, is equal to 10.4 ms for 128-byte EF
packet size and equal to 18.7 ms for 1518-byte EF packet size.

These results are similar to the experimental results
in~\cite{ferrari2000}. The more the EF packet size increases, the
more significant values on x axis become dense. This behavior is
described in~\cite{ferrari2000}. Another interesting phenomena is
that, in both tests, the WFQ profile moves to higher delays faster
than the PQ profile. This is because the deployed schedulers have
different factors affecting the queuing delay. With the PQ
scheduler, the queuing delay of priority packets is affected only by
the ongoing transmission of lower priority packets. Conversely, with
the WFQ scheduler (SCFQ implementation), the factors are clearly
described in~\cite{stiliadis1996}. The maximum latency introduced by
this scheduler is:
\begin{equation}
  Latency = \frac{L_i}{\rho_i} + \frac{L_{max}}{r} (V - 1)
\end{equation}
in which $L_i$ is the maximum packet size of session $i$, $L_{max}$
is the maximum packet size among all sessions, $\rho_i$ is the
allocated rate to session $i$, $r$ is the total service rate and $V$
is the number of active sessions. Hence, queue latency due to the
WFQ scheduler increases with the packet size and with the number of
active queues/sessions (the latter has been positively verified
through simulations). Studying the scheduler algorithm and matching
it against the traffic characteristics is clear that all the
background packets that enter the queue in between two EF packet
arrivals, will be scheduled and serviced up to the allocated service
time. Only after the transmission of these packets, the new EF
arrival will be served. This explains the WFQ behavior against the
PQ scheduler.

\begin{figure}[t]
  \centering
  \includegraphics[width=0.85\linewidth]{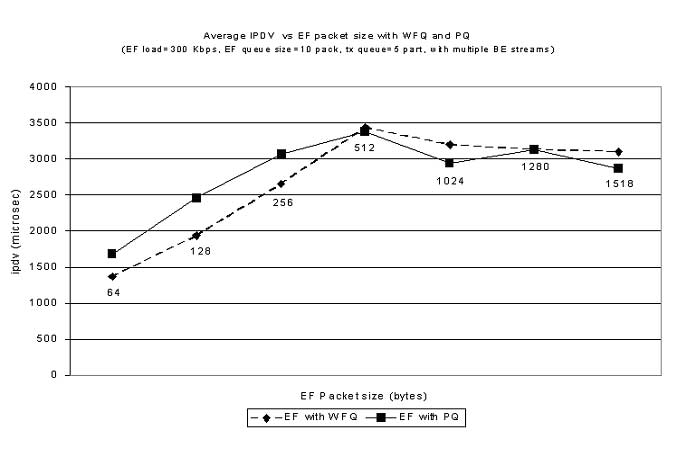}
  \caption{Average IPDV with WFQ and PQ for different EF packet sizes --
           experiment~\cite{ferrari2000}}
  \label{fig:11}
\end{figure}

\begin{figure}[t]
  \centering
  \includegraphics[width=0.85\linewidth]{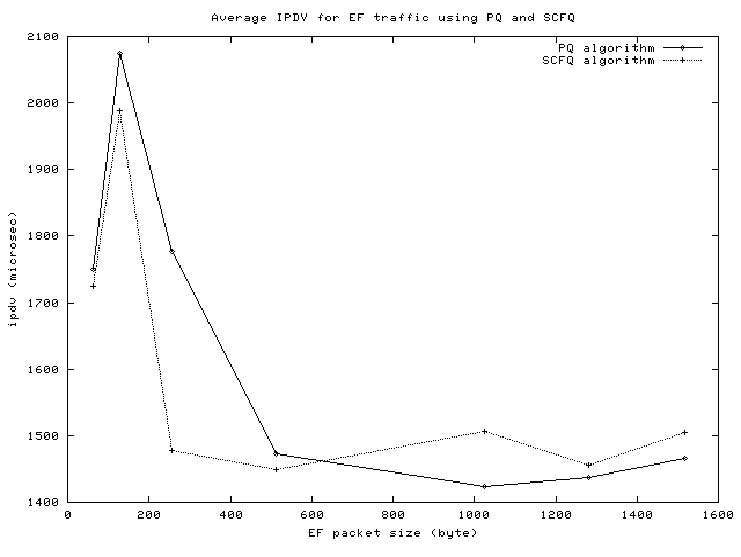}
  \caption{Average IPDV with WFQ and PQ for different EF packet sizes --
           simulation}
  \label{fig:12}
\end{figure}

\begin{figure}[t]
  \centering
  \includegraphics[width=0.75\linewidth]{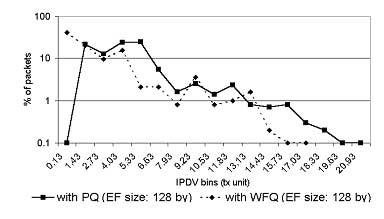}
  \caption{Frequency distributed IPDV with WFQ and PQ for 128-byte EF
           packet size -- experiment~\cite{ferrari2000}}
  \label{fig:13}
\end{figure}

\begin{figure}[t]
  \centering
  \includegraphics[width=0.85\linewidth]{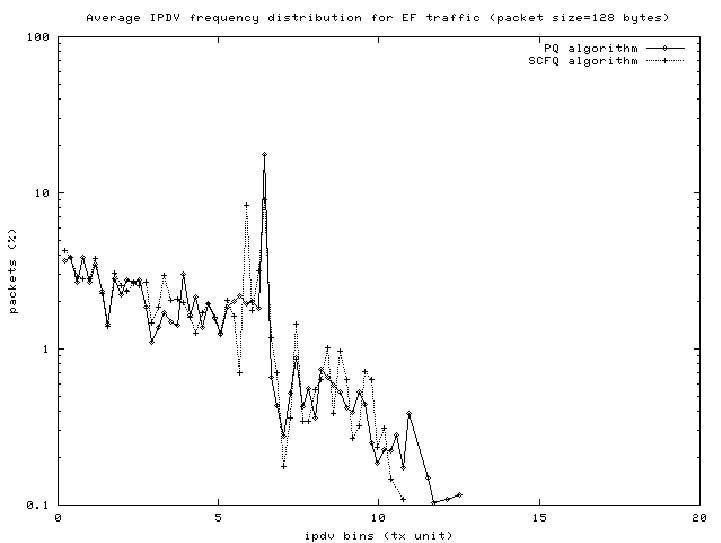}
  \caption{Frequency distributed IPDV with WFQ and PQ for 128-byte EF
           packet size -- simulation}
  \label{fig:14}
\end{figure}

\begin{figure}[t]
  \centering
  \includegraphics[width=0.75\linewidth]{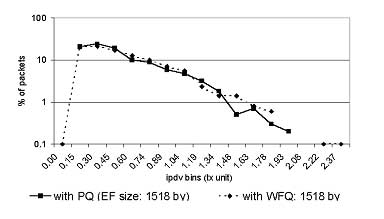}
  \caption{Frequency distributed IPDV with WFQ and PQ for 1518-byte EF
           packet size -- experiment~\cite{ferrari2000}}
  \label{fig:15}
\end{figure}

\begin{figure}[t]
  \centering
  \includegraphics[width=0.85\linewidth]{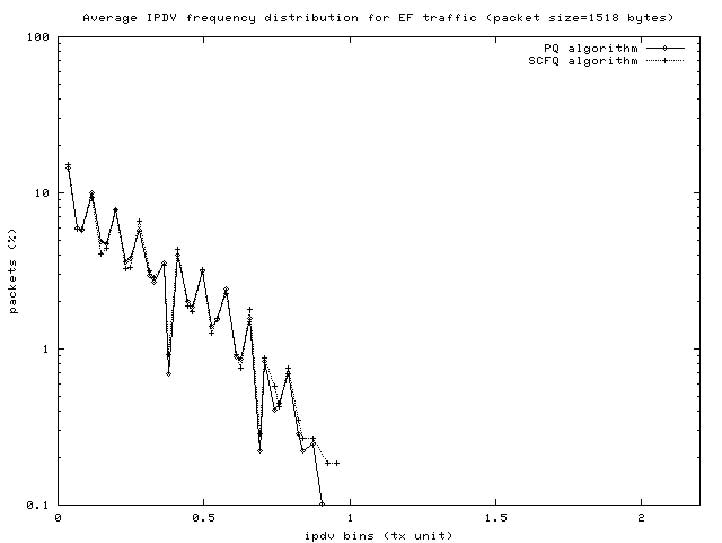}
  \caption{Frequency distributed IPDV with WFQ and PQ for 1518-byte EF
           packet size -- simulation}
  \label{fig:16}
\end{figure}

Now, the IPDV analysis is approached (Figures~\ref{fig:11}
and~\ref{fig:12}). In these diagrams the trend is different, however
it is interesting to notice that curves intersect for similar EF
packet size values. For the IPDV parameter, simulations apparently
are not useful to track the network behavior. But looking at
Figures~\ref{fig:13},~\ref{fig:14},~\ref{fig:15} and~\ref{fig:16},
it is possible to assert that worthy information can be gathered.
These diagrams show the IPDV frequency distribution for both real
and simulated test beds and for 128-byte and 1518-byte EF packet
sizes. Measurements of IPDV is given as multiple of Transmission
Units, that is the transmission time at line rate for the reference
EF packet.

These diagrams show that for the IPDV parameter, no sensitive
differences are detected between the two schedulers.

\section{Summary and conclusions}

Through all this work it has been showed that simulations are a
useful activity for advanced networking studies. In particular for
the Differentiated Services architecture, an experimental test bed
and related tests have been replicated in the Network Simulator.
Such activity needed a re-engineering of the NS DiffServ module and
valuable contributes have been made in term of richness of
functional blocks and performance monitoring capabilities.

Results show that PQ and WFQ schedulers behaviors are properly
described even if deployed in a complex scenario such is the
experimental test bed. Moreover, it has been pointed out that with
the correct assumptions, the high complex real network has been
meaningfully described by a simpler model. Both average and
frequency distribution of OWD parameter can be considered useful for
the description of traffic behavior in different scenarios.
Conversely, only the frequency distribution of the IPDV provided
meaningful information.

\section*{Acknowledgments}

The author wishes to thank Kari Heikkinen who led this work and
Tiziana Ferrari who provided worthy critical comments and beneficial
suggestions.

\FloatBarrier

\bibliographystyle{unsrt}
\bibliography{preprint}

\end{document}